\documentclass[runningheads]{llncs}
\usepackage{graphicx}
\usepackage[utf8]{inputenc} 
\usepackage[T1]{fontenc}    
\usepackage{hyperref}       
\usepackage{url}            
\usepackage{booktabs}       
\usepackage{amsfonts}       
\usepackage{nicefrac}       
\usepackage{microtype}      
\usepackage{comment}
\usepackage{amsmath}
\usepackage{amssymb}
\usepackage{xcolor}
\usepackage{booktabs}

\usepackage{caption, subcaption}

\usepackage[margin=1in]{geometry}
\usepackage[sort&compress,numbers]{natbib}
\usepackage{xr}
\usepackage{subfiles}
\usepackage{cleveref}

\newcommand{\x}{\mathbf{x}}

\newcommand{\K}{\mathbf{k}}

\newcommand{\ft}{\mathcal{F}}
\newcommand{\ift}{\mathcal{F}^{-1}}
\newcommand{\R}{\mathcal{R}}
\newcommand{\E}{\mathcal{E}}

\newcommand{\C}{\mathbb{C}}

\newcommand{\norm}[1]{\left\lVert#1\right\rVert}

\DeclareMathOperator{\argmin}{argmin}
\DeclareMathOperator{\ssim}{SSIM}
\DeclareMathOperator{\CNN}{CNN}

\DeclareMathOperator{\vn}{VN}
\DeclareMathOperator{\vnu}{VNU}

\DeclareMathOperator{\vnuk}{VNU-K}
\DeclareMathOperator{\e2evn}{E2E-VN}

\begin{document}
\title{End-to-End Variational Networks for Accelerated MRI Reconstruction}
%
%

\author{
Anuroop Sriram\inst{1}\thanks{Equal contribution} \and
Jure Zbontar\inst{1}$^\star$ \and
Tullie Murrell\inst{1} \and
Aaron Defazio\inst{1} \and
C. Lawrence Zitnick\inst{1} \and
Nafissa Yakubova\inst{1} \and
Florian Knoll\inst{2} \and
Patricia Johnson\inst{2}
}
\institute{
Facebook AI Research (FAIR) \and
NYU School of Medicine
}
\authorrunning{A. Sriram et al.}

%
\maketitle              
\begin{abstract}
The slow acquisition speed of magnetic resonance imaging (MRI) has led to the development of two complementary methods: acquiring multiple views of the anatomy simultaneously (parallel imaging) and acquiring fewer samples than necessary for traditional signal processing methods (compressed sensing). While the combination of these methods has the potential to allow much faster scan times, reconstruction from such undersampled multi-coil data has remained an open problem. In this paper, we present a new approach to this problem that extends previously proposed variational methods by learning fully end-to-end. Our method obtains new state-of-the-art results on the fastMRI dataset \cite{zbontar2018fastmri} for both brain and knee MRIs.\footnote{Code is available at \url{https://github.com/facebookresearch/fastMRI}.}

\end{abstract}

\section{Introduction}
Magnetic Resonance Imaging (MRI) is a powerful diagnostic tool for a  variety of disorders, but its utility is often limited by its slow speed compared to competing modalities like CT or X-Rays. Reducing the time required for a scan would decrease the cost of MR imaging, and allow for acquiring scans in situations where a patient cannot stay still for the current minimum scan duration.

One approach to accelerating MRI acquisition, called Parallel Imaging (PI) \cite{sodickson1997simultaneous,pruessmann1999sense,griswold2002Grappa}, utilizes multiple receiver coils to simultaneously acquire multiple views of the underlying anatomy, which are then combined in software. Multi-coil imaging is widely used in current clinical practice. A complementary approach to accelerating MRIs acquires only a subset of measurements and utilizes Compressed Sensing (CS) \cite{candes2006compressive,Lustig2007} methods to reconstruct the final image from these undersampled measurements. The combination of PI and CS, which involves acquiring undersampled measurements from multiple views of the anatomy, has the potential to allow faster scans than is possible by either method alone. Reconstructing MRIs from such undersampled multi-coil measurements has remained an active area of research.

MRI reconstruction can be viewed as an inverse problem and previous research has proposed neural networks whose design is inspired by the optimization procedure to solve such a problem \cite{varnet,putzky2019rim,putzky2019invert,liang2019deep}. A limitation of such an approach is that it assumes the forward process is completely known, which is an unrealistic assumption for the multi-coil reconstruction problem. In this paper, we present a novel technique for reconstructing MRI images from undersampled multi-coil data that does not make this assumption. We extend previously proposed variational methods by learning the forward process in conjunction with reconstruction, alleviating this limitation. We show through experiments on the fastMRI dataset that such an approach yields higher fidelity reconstructions.

Our contributions are as follows: 1) we extend the previously proposed variational network model by learning completely end-to-end; 2) we explore the design space for the variational networks to determine the optimal intermediate representations and neural network architectures for better reconstruction quality; and 3) we perform extensive experiments using our model on the fastMRI dataset and obtain new state-of-the-art results for both the knee and the brain MRIs.

\section{Background and Related Work}

\subsection{Accelerated MRI acquisition}\label{sec:accel_mri}

An MR scanner images a patient's anatomy by acquiring measurements in the frequency domain, called \emph{k-space}, using a measuring instrument called a receiver coil. 
The image can then be obtained by applying an inverse multidimensional Fourier transform $\ift$ to the measured k-space samples.
The underlying image $\x \in \C^M$ is related to the measured k-space samples $\K \in \C^M$ as
\begin{equation}
    \K = \ft(\x) + \epsilon,
\end{equation}
where $\epsilon$ is the measurement noise and $\ft$ is the fourier transform operator.

Most modern scanners contain multiple receiver coils. Each coil acquires k-space samples that are modulated by the sensitivity of the coil to the MR signal arising from different regions of the anatomy. Thus, the $i$-th coil measures:
\begin{equation}
    \K_i = \ft(S_i\x) + \epsilon_i, i = 1, 2, \dots, N,
\end{equation}
where $S_i$ is a complex-valued diagonal matrix encoding the position dependent sensitivity map of the $i$-th coil and $N$ is the number of coils. The sensitivity maps are normalized to satisfy~\cite{uecker2014espirit}:
\begin{equation}\label{eq:sens_rss}
\sum_{i=1}^{N} S_i^* S_i = 1
\end{equation}

The speed of MRI acquisition is limited by the number of k-space samples obtained. This acquisition process can be accelerated by obtaining undersampled k-space data, $\tilde{\K_i} = M \K_i$,
where $M$ is a binary mask operator that selects a subset of the k-space points and $\tilde{\K_i}$ denotes the measured k-space data. The same mask is used for all coils. Applying an inverse Fourier transform naively to this under-sampled k-space data results in aliasing artifacts.

Parallel Imaging can be used to accelerate imaging by exploiting redundancies in k-space samples measured by different coils. The sensitivity maps $S_i$ can be estimated using the central region of k-space corresponding to low frequencies, called the \emph{Auto-Calibration Signal (ACS)}, which is typically fully sampled. To accurately estimate these sensitivity maps, the ACS must be sufficiently large, which limits the maximum possible acceleration.

\subsection{Compressed Sensing for Parallel MRI Reconstruction}\label{sec:cs_mri}

Compressed Sensing \cite{donoho2006compressed} enables reconstruction of images by using fewer k-space measurements than is possible with classical signal processing methods by enforcing suitable priors. 
Classical compressed sensing methods solve the following optimization problem:
\begin{align}
    \hat{\x} &= \argmin_{\x} \frac{1}{2} \sum_{i} \norm{M \ft(S_i \x) - \tilde{\K_i}}^2 + \lambda \Psi(\x) \label{eq:pics_mri1}\\
        &= \argmin_{\x} \frac{1}{2} \norm{A(\x) - \tilde{\K}}^2 + \lambda \Psi(\x) \label{eq:pics_mri2},
\end{align}
where $\Psi$ is a regularization function that enforces a sparsity constraint,
$A$ is the linear forward operator that multiplies by the sensitivity maps, applies 2D fourier transform and then under-samples the data, and $\tilde{\K}$ is the vector of masked k-space data from all coils. 
This problem can be solved by iterative gradient descent methods. In the $t$-th step the image is updated from $\x^t$ to $\x^{t+1}$ using:
\begin{equation}\label{eq:pics_mri_gd}
    \x^{t+1} = \x^t - \eta^t \left( A^* (A(\x) - \tilde{\K}) + \lambda \Phi(\x^t) \right),
\end{equation}
where $\eta^t$ is the learning rate, $\Phi(\x)$ is the gradient of $\Psi$ with respect to $\x$, and $A^*$ is the hermitian of the forward operator $A$.

\subsection{Deep Learning for Parallel MRI Reconstruction}

In the past few years, there has been rapid development of deep learning based approaches to MRI reconstruction~\citep{varnet,putzky2019invert,putzky2019rim,Knoll2019DeepLM,liang2019deep,SchlemperCHPR17_dynamic}. A comprehensive survey of recent developments in using deep learning for parallel MRI reconstruction can be found in \cite{Knoll2019DeepLM}. Our work builds upon the Variational Network (VarNet)~\cite{varnet}, which consists of multiple layers, each modeled after a single gradient update step in equation \ref{eq:pics_mri_gd}. Thus, the $t$-th layer of the VarNet takes $\x^{t}$ as input and computes $\x^{t+1}$ using:
\begin{equation} \label{eq:vn_layer}
    \x^{t+1} = \x^{t} - \eta^t A^* (A(\x^{t}) - \tilde{\K}) + \CNN(\x^{t}),
\end{equation}
where $\CNN$ is a small convolutional neural network that maps complex-valued images to complex-valued images of the same shape.
The $\eta^t$ values as well as the parameters of the $\CNN$s are learned from data.

The $A$ and $A^*$ operators involve the use of sensitivity maps which are computed using a traditional PI method and fed in as additional inputs. As noted in section \ref{sec:accel_mri}, these sensitivity maps cannot be estimated accurately when the number of auto-calibration lines is small, which is necessary to achieve higher acceleration factors. As a result, the performance of the VarNet degrades significantly at higher accelerations. We alleviate this problem in our model by learning to predict the sensitivity maps from data as part of the network.

\section{End-to-End Variational Network}\label{sec:e2e-vn}

\begin{figure}
    \centering
    \includegraphics[scale=0.7]{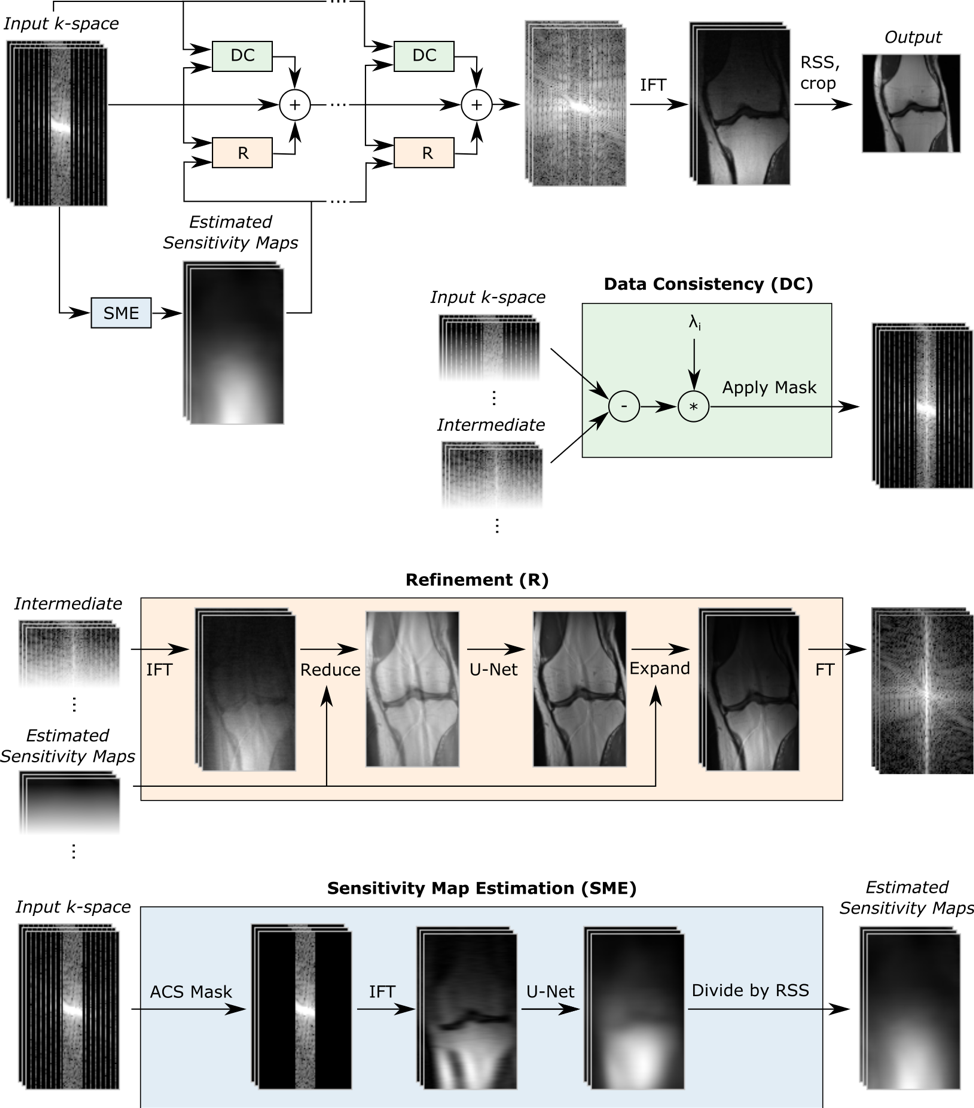}
    \caption{\textbf{Top}: Block diagram of our model which takes under-sampled k-space as input and applies several cascades, followed by an inverse Fourier transform (IFT) and an RSS transform. The \textbf{Data Consistency (DC)} module computes a correction map that brings the intermediate k-space closer to the measured k-space values. The \textbf{Refinement (R)} module maps multi-coil k-space data into one image, applies a U-Net, and then back to multi-coil k-space data. The \textbf{Sensitivity Map Estimation (SME)} module estimates the sensitivity maps used in the Refinement module.}
    \label{fig:vnmodel}
\end{figure}

Let $\K_0 = \tilde{\K}$ be the vector of masked multi-coil k-space data. 
Similar to the VarNet, our model takes this masked k-space data $\K_0$ as input and applies a number of refinement steps of the same form. We refer to each of these steps as a \emph{cascade} (following \cite{SchlemperCHPR17_dynamic}), to avoid overloading the term "layer" which is already heavily used. Unlike the VN, however, our model uses k-space intermediate quantities rather than image-space quantities. We call the resulting method the End-to-End Variational Network or E2E-VarNet.

\subsection{Preliminaries}
To simplify notation, we first define two operators: the expand operator ($\E$) and the reduce operator ($\R$). The \emph{expand} operator ($\E$) takes the image $\x$ and sensitivity maps as input and computes the corresponding image seen by each coil in the idealized noise-free case:
\begin{equation}
    \E(\x) = (\x_1, ..., \x_N) = (S_1 \x, ..., S_N \x).
\end{equation}
where $S_i$ is the sensitivity map of coil $i$. We do not explicitly represent the sensitivity maps as inputs for the sake of readability. The inverse operator, called the \emph{reduce} operator ($\R$) combines the individual coil images:
\begin{equation}
    \R(\x_1, ... \x_N) = \sum_{i=1}^{N} S_i^* \x_i.
\end{equation}



Using the expand and reduce operators, $A$ and $A^*$ can be written succinctly as $A = M \circ \ft \circ \E$ and $A^* = \R \circ \ift \circ M$.

\subsection{Cascades}
Each cascade in our model applies a refinement step similar to the gradient descent step in equation \ref{eq:vn_layer}, except that the intermediate quantities are in k-space. Applying $\ft \circ \E$ to both sides of \ref{eq:vn_layer} gives the corresponding update equation in k-space:
\begin{equation} \label{eq:e2evn-update}
    \K^{t+1} = \K^t - \eta^t M (\K^t - \tilde{\K}) + G(\K^t)
\end{equation}
 where $G$ is the \emph{refinement module} given by:
\begin{equation} \label{eq:refinement}
G(\K^t) = \ft \circ \E \circ \CNN (\R \circ \ift (\K^t)).
\end{equation}
Here, we use the fact that $\x^t = \R \circ \ift (\K^t$). $\CNN$ can be any parametric function that takes a complex image as input and maps it to another complex image. Since the CNN is applied after combining all coils into a single complex image, the same network can be used for MRIs with different number of coils.

Each cascade applies the function represented by equation \ref{eq:e2evn-update} to refine the k-space. In our experiments, we use a U-Net \cite{Ronneberger2015unet} for the $\CNN$.

\subsection{Learned sensitivity maps}
The expand and reduce operators in equation \ref{eq:refinement} take sensitivity maps $(S_1, ..., S_N)$ as inputs. In the original VarNet model, these sensitivity maps are computed using the ESPIRiT algorithm \cite{uecker2014espirit} and fed in to the model as additional inputs. In our model, however, we estimate the sensitivity maps as part of the reconstruction network using a \emph{Sensitivity Map Estimation (SME)} module:
\begin{equation}
    H = \text{dSS} \circ \CNN \circ \ift \circ M_{\text{center}}.
\end{equation}

The $M_{\text{center}}$ operator zeros out all lines except for the autocalibration or ACS lines (described in Section \ref{sec:accel_mri}). This is similar to classical parallel imaging approaches which estimate sensitivity maps from the ACS lines. The CNN follows the same architecture as the CNN in the cascades, except with fewer channels and thus fewer parameters in intermediate layers. This CNN is applied to each coil image independently.
Finally, the $\text{dSS}$ operator normalizes the estimated sensitivity maps to ensure that the property in equation \ref{eq:sens_rss} is satisfied.

\subsection{E2E-VarNet model architecture}

As previously described, our model takes the masked multi-coil k-space $\K_0$ as input. First, we apply the SME module to $\K_0$ to compute the sensitivity maps. Next we apply a series of cascades, each of which applies the function in equation \ref{eq:e2evn-update}, to the input k-space to obtain the final k-space representation $K^T$. This final k-space representation is converted to image space by applying an inverse Fourier transform followed by a root-sum-squares (RSS) reduction for each pixel:
\begin{equation}
    \hat{\x} = RSS(\x_1, ..., \x_N) = \sqrt{\sum_{i=1}^{N} |\x_i| ^ 2}
\end{equation}
where $\x_i = \ift(K^T_i)$ and $K^T_i$ is the k-space representation for coil $i$. The model is illustrated in figure \ref{fig:vnmodel}.

All of the parameters of the network, including the parameters of the CNN model in SME, the parameters of the CNN in each cascade along with the $\eta^t$s, are estimated from the training data by minimizing the structural similarity loss, $ J(\hat{\x}, \x^*) = - \ssim(\hat{\x}, \x^*)$,
where SSIM is the Structural Similarity index~\cite{wang2003multiscale} and $\hat{\x}$, $\x^*$ are the reconstruction and ground truth images, respectively.


\section{Experiments}

\subsection{Experimental setup}

We designed and validated our method using the multicoil track of the fastMRI dataset \cite{zbontar2018fastmri} which is a large and open dataset of knee and brain MRIs. To validate the various design choices we made, we evaluated the following models on the knee dataset: 
\begin{enumerate}
    \item Variational network \cite{varnet} ($\vn$)
    \item Variational network with the shallow CNNs replaced with U-Nets ($\vnu$)
    \item Similar to $\vnu$, but with k-space intermediate quantities ($\vnuk$)
    \item Our proposed end-to-end variational network model ($\e2evn$)
\end{enumerate}

The $\vn$ model employs shallow convolutional networks with RBF kernels that have about 150K parameters in total. $\vnu$ replaces these shallow networks with U-Nets to ensure a fair comparison with our model. $\vnuk$ is similar to our proposed model but uses fixed sensitivity maps computed using classical parallel imaging methods. The difference in reconstruction quality between $\vnu$ and $\vnuk$ shows the value of using k-space intermediate quantities for reconstruction, while the difference between $\vnuk$ and $\e2evn$ shows the importance of learning sensitivity maps as part of the network.

We used the same model architecture and training procedure for the $\vn$ model as in the original VarNet \cite{varnet} paper. For each of the other models, we used $T = 12$ cascades, containing a total of about 29.5M parameters. The $\e2evn$ model contained an additional 0.5M parameters in the SME module, taking the total number of parameters to 30M. We trained these models using the Adam optimizer with a learning rate of 0.0003 for 50 epochs, without using any regularization or data augmentation techniques.

We used two types of under-sampling masks: \emph{equispaced masks} $M_e(r, l)$, which sample $l$ low-frequency lines from the center of k-space and every $r$-th line from the remaining k-space; and \emph{random masks} $M_r(a, f)$, which sample a fraction $f$ of the full width of k-space for the ACS lines in addition to a subset of higher frequency lines, selected uniformly at random, to make the overall acceleration equal to $a$. These random masks are identical to those used in \cite{zbontar2018fastmri}. We also use equispaced masks as they are easier to implement in MRI machines.

\vspace{5mm}
\begin{figure}
    \centering
    \includegraphics[width=0.85\linewidth]{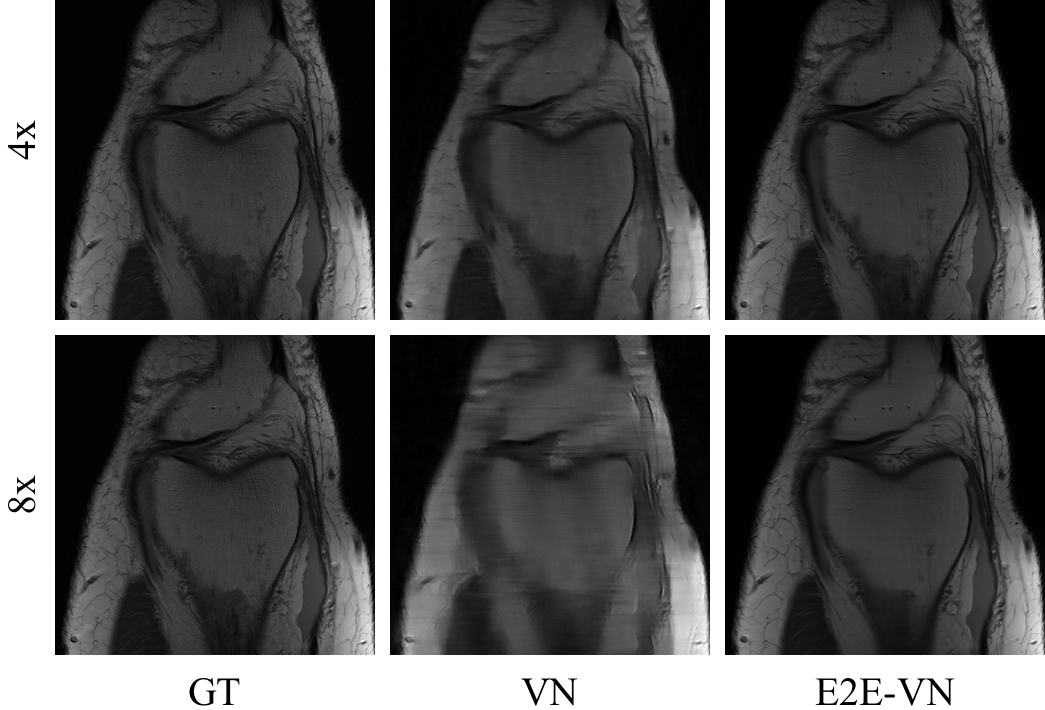}
    \caption{Examples comparing the ground truth (GT) to VarNet (VN) and E2E-VarNet (E2E-VN). At $8\times$ acceleration, the VN images contain severe artifacts.}
    \label{fig:recons}
\end{figure}

\subsection{Results}
Tables \ref{tab:result_equispaced} and \ref{tab:result_random} show the results of our experiments for equispaced and random masks respectively, over a range of down-sampling mask parameters.
The $\vnu$ model outperforms the baseline $\vn$ model by a large margin due to its larger capacity and the multi-scale modeling ability of the U-Nets. $\vnuk$ outperforms $\vnu$ demonstrating the value of using k-space intermediate quantities. $\e2evn$ outperforms $\vnuk$ showing the importance of learning sensitivity maps as part of the network. It is worth noting that the relative performance does not depend on the type of mask or the mask parameters. Some example reconstructions are shown in figure \ref{fig:recons}.

\begin{table}[htb]
\parbox{.47\linewidth}{
\centering
    \begin{tabular}{c c c c}
            \toprule
            Accel($r$) & Num ACS($l$) & Model & SSIM \\
            \midrule
      &   & $\vn$ & 0.818 \\ 
    4 & 30 & $\vnu$ & 0.919 \\ 
      &  & $\vnuk$ & 0.922 \\ 
      &   & $\e2evn$ & \textbf{0.923} \\
    \midrule 
    
      &   & $\vn$ & 0.795 \\
    6 &22 & $\vnu$ & 0.904 \\ 
      &  & $\vnuk$ & 0.907 \\ 
      &   & $\e2evn$ & \textbf{0.907} \\
    \midrule 
    
      &   & $\vn$ & 0.782 \\ 
    8 & 16& $\vnu$ & 0.889 \\ 
      & & $\vnuk$ & 0.891 \\ 
      &   & $\e2evn$ & \textbf{0.893} \\
    \bottomrule
    \end{tabular} 
    \caption{Experimental results using equispaced masks on the knee MRIs \label{tab:result_equispaced}}
}
\hfill
\parbox{.47\linewidth}{
\centering
    \begin{tabular}{c c c c}
            \toprule
            Accel($a$) & Frac ACS($f$) & Model & SSIM \\
            \midrule
      &   & $\vn$ & 0.813 \\ 
    4 & 0.08  & $\vnu$ & 0.906 \\ 
      & & $\vnuk$ & 0.907 \\ 
      &   & $\e2evn$ & \textbf{0.910} \\
    \midrule 
    
      &   & $\vn$ & 0.767 \\ 
    6 & 0.06  & $\vnu$ & 0.886 \\ 
      &   & $\vnuk$ & 0.887 \\ 
      &   & $\e2evn$ & \textbf{0.892} \\
    \midrule 
    
      &   & $\vn$ & 0.762 \\ 
    8 & 0.04  & $\vnu$ & 0.870 \\ 
      &   & $\vnuk$ & 0.871 \\ 
      &   & $\e2evn$ & \textbf{0.878} \\
    \bottomrule 
    \end{tabular} 
    \caption{Experimental results using random masks on the knee MRIs \label{tab:result_random}}
}
\end{table}

\subsubsection{Significance of learning sensitivity maps} 
Figure \ref{fig:eq_mask_params} shows the SSIM values for each model with various equispaced mask parameters. In all cases, learning the sensitivity maps improves the SSIM score. Notably, this improvement in SSIM is larger when the number of low frequency lines is smaller. As previously stated, the quality of the estimated sensitivity maps tends to be poor when there are few ACS lines, which leads to a degradation in the final reconstruction quality. The $\e2evn$ model is able to overcome this limitation and generate good reconstructions even with a small number of ACS lines.

\begin{figure}
\centering
    \parbox{0.47\linewidth}{
        \centering
        \includegraphics[scale=0.5]{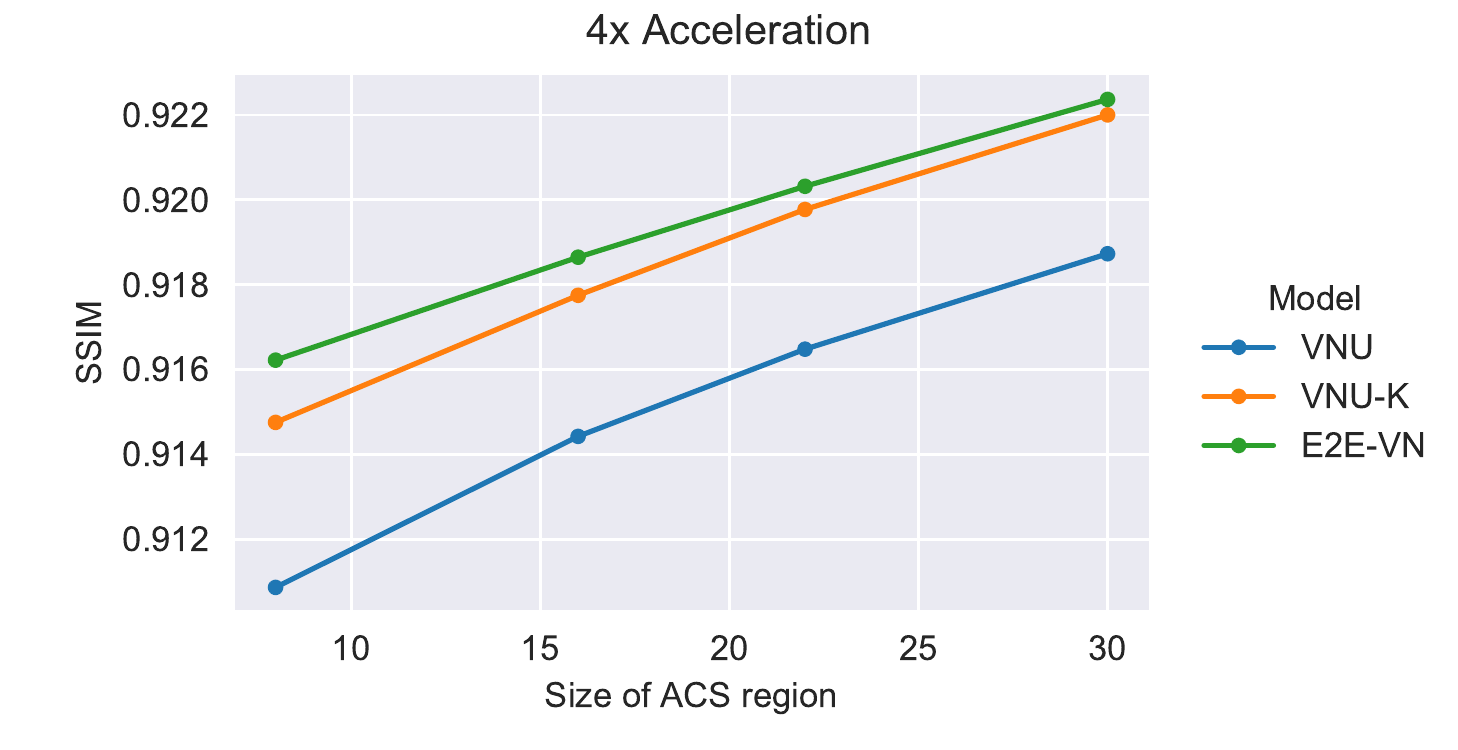}
    }
    \parbox{0.5\linewidth}{
        \centering
        \includegraphics[scale=0.5]{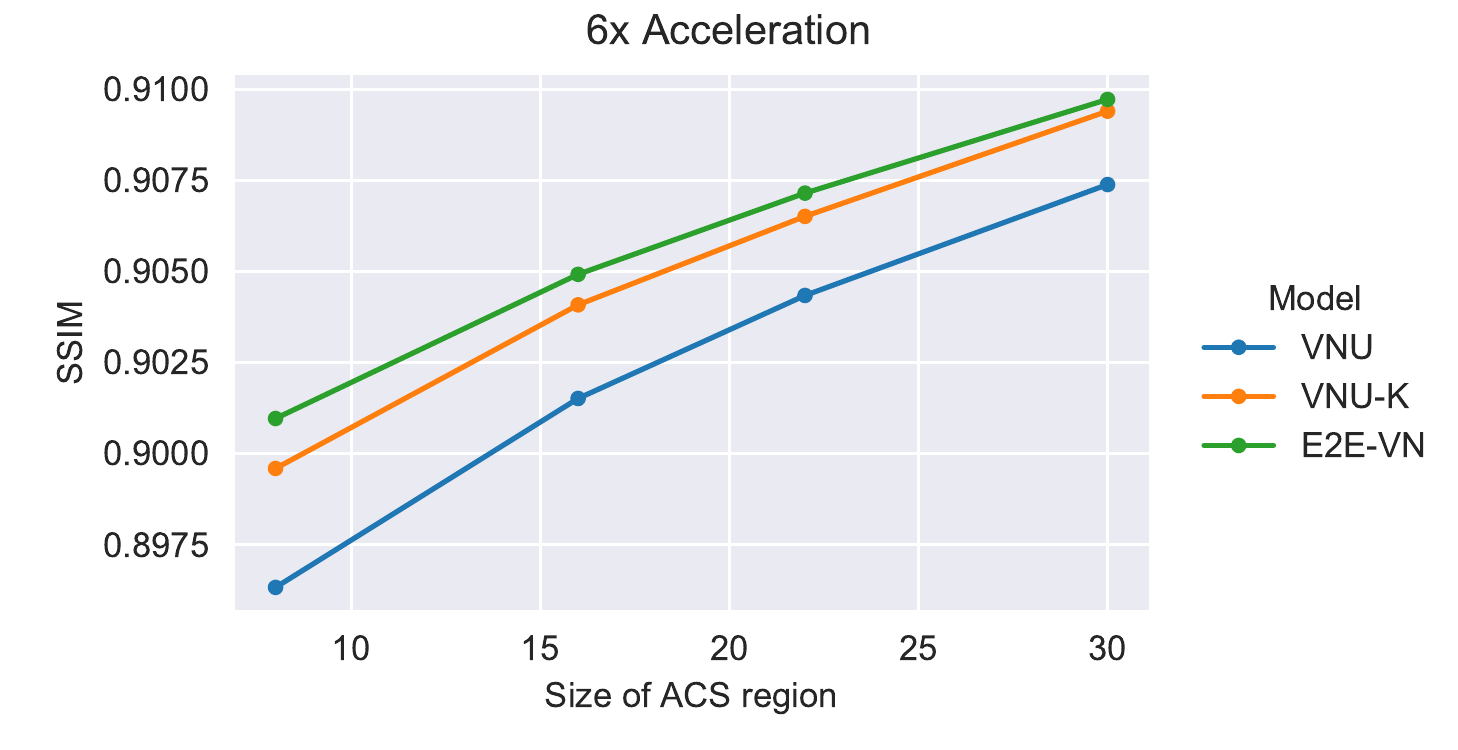}
    }
    \caption{Effect of different equispaced mask parameters on reconstruction quality at 4$\times$ and 6$\times$ acceleration.
    }
    \label{fig:eq_mask_params}
\end{figure}

\subsubsection{Experiments on test data}


\begin{table}[htb]
\centering
    \begin{tabular}{c c c c c c c c}
        \toprule
        Dataset & Model & \multicolumn{3}{c}{4$\times$ Acceleration} & \multicolumn{3}{c}{8$\times$ Acceleration} \\
         \cmidrule(r){3-5} \cmidrule(r){6-8}
             & & SSIM & NMSE & PSNR & SSIM & NMSE & PSNR \\
        \midrule
                 & E2E-VN & \textbf{0.930} & 0.005 & 40 & \textbf{0.890} & 0.009 & 37 \\
                 & SubtleMR & 0.929 & 0.005 & 40 & 0.888 & 0.009 & 37 \\
                Knee & AIRS Medical & 0.929 & 0.005 & 40 & 0.888 & 0.009 & 37 \\
                 & SigmaNet & 0.928 & 0.005 & 40 & 0.888 & 0.009 & 37 \\
                 & i-RIM & 0.928 & 0.005 & 40 & 0.888 & 0.009 & 37 \\
        \midrule
        Brain  & E2E-VN & \textbf{0.959} & 0.004 & 41& \textbf{0.943} & 0.008 & 38 \\
                         & U-Net & 0.945 & 0.011 & 38 &  0.915 & 0.023 & 35 \\
        \bottomrule
    \end{tabular}
    \caption{Results on the test data compared with the best models on the fastMRI leaderboard}
    \label{tab:test_res}
\end{table}

Table \ref{tab:test_res} shows our results on the test datasets for both the brain and knee MRIs compared with the best models on the fastMRI leaderboard\footnote{\url{http://fastmri.org/leaderboards}}. To obtain these results, we used the same training procedure as our previous experiments, except that we trained on both the training and validation sets for 100 epochs. We used the same type of masks that are used for the fastMRI paper~\cite{varnet}. Our model outperforms all other models published on the fastMRI leaderboard for both anatomies.


\section{Conclusion}
In this paper, we introduced End-to-End Variational Networks for multi-coil MRI reconstruction. While MRI reconstruction can be posed as an inverse problem, multi-coil MRI reconstruction is particularly challenging because the forward process (which is determined by the sensitivity maps) is not completely known. We alleviate this problem by estimating the sensitivity maps within the network, and learning fully end-to-end. Further, we explored the architecture space to identify the best neural network layers and intermediate representation for this problem, which allowed our model to obtain new state-of-the art results on both brain and knee MRIs.

The quantitative measures we have used only provide a rough estimate for the quality of the reconstructions. Many clinically important details tend to be subtle and limited to small regions of the MR image. Rigorous clinical validation needs to be performed before such methods can be used in clinical practice to ensure that there is no degradation in the quality of diagnosis.

\bibliographystyle{plainnat}
\bibliography{vn}

\section{Supplementary Materials}




\subsection{Dithering as post-processing}

The Structural Similarity (SSIM) loss \cite{wang2004image}  we used to train our models has a tendency to produce overly smooth reconstructions even when all of the diagnostic content is preserved. We noticed a similar behavior with other frequently used loss functions like mean squared error, mean absolute error, etc. Sriram et al. \cite{sriram2019grappanet} found that dithering the image by adding a small amount of random gaussian noise helped enhance the perceived sharpness of their reconstructions. We found that the same kind of dithering helped improve the sharpness of our reconstructions, but we tuned the scale of the noise by manual inspection.

Similar to \cite{sriram2019grappanet}, we adjusted the scale of the added noise to the brightness of the image around each pixel to avoid obscuring dark areas of the reconstruction. Specifically, we first normalize the image by dividing each pixel by the maximum pixel intensity. Then we add zero-mean random gaussian noise to each pixel. The standard deviation of the noise at a given pixel location is equal to $\sigma$ times the square root of the local median computed over a patch of $11\times11$ pixels around that pixel location. We set $\sigma = 0.02$ for the brain images and non fat suppressed knee images, and $\sigma = 0.03$ for the fat suppressed knee images.

Example reconstructions with and without noise are shown in 
\cref{fig:brain4x,fig:brain8x,fig:knee4x,fig:knee8x}. The dithered images look more natural, especially at higher accelerations.






\begin{figure}[ht]
\begin{subfigure}{\textwidth}
    \centering
    \includegraphics[width=0.32\textwidth]{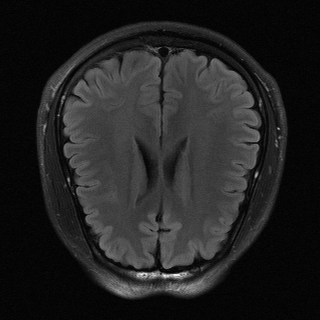}
    \includegraphics[width=0.32\textwidth]{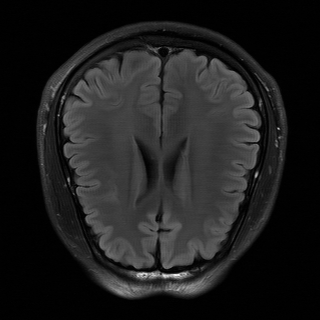}
    \includegraphics[width=0.32\textwidth]{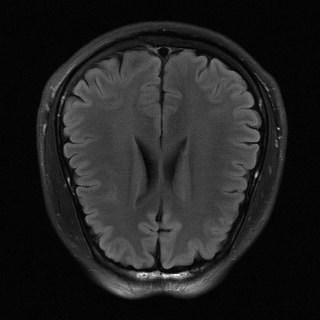}
\end{subfigure}

\vspace{5 mm}

\begin{subfigure}{\textwidth}
    \centering
    \includegraphics[width=0.32\textwidth]{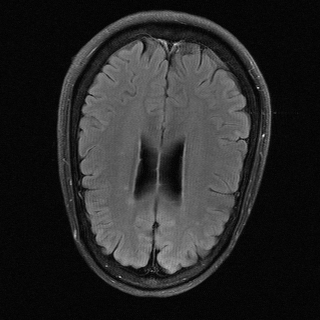}
    \includegraphics[width=0.32\textwidth]{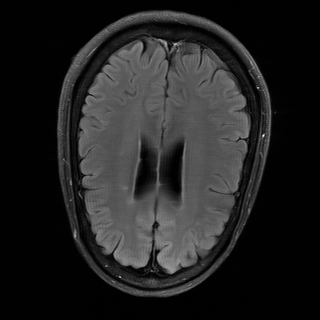}
    \includegraphics[width=0.32\textwidth]{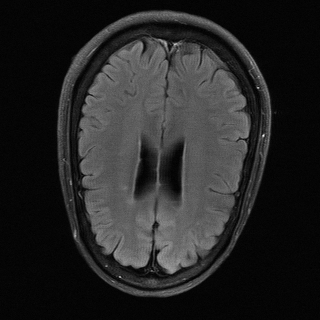}
\end{subfigure}
\caption[*]{Brain images showing ground truth (left), reconstruction (middle) and reconstruction with added noise (right) at $4\times$ acceleration}
\label{fig:brain4x}
\end{figure}

\begin{figure}[ht]
\begin{subfigure}{\textwidth}
    \centering
    \includegraphics[width=0.32\textwidth]{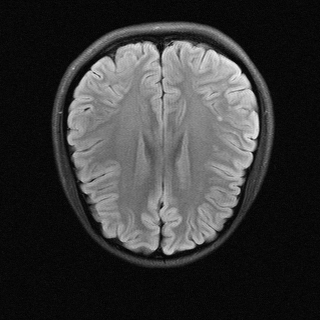}
    \includegraphics[width=0.32\textwidth]{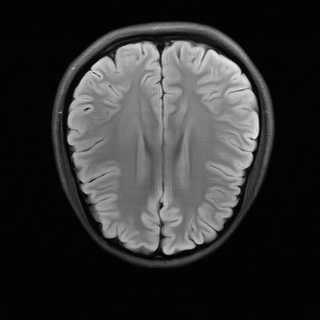}
    \includegraphics[width=0.32\textwidth]{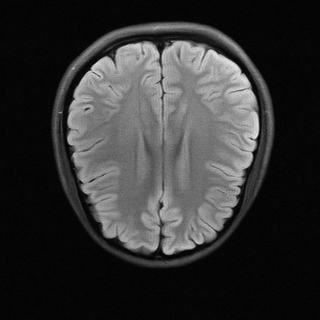}
\end{subfigure}

\vspace{5 mm}

\begin{subfigure}{\textwidth}
    \centering
    \includegraphics[width=0.32\textwidth]{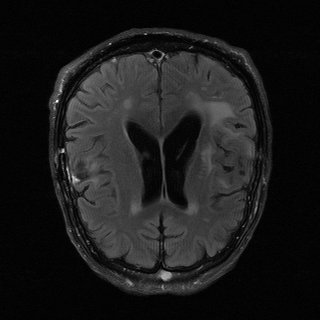}
    \includegraphics[width=0.32\textwidth]{figs/additional/brain_8x_file_brain_AXFLAIR_201_6002990_gt.png}
    \includegraphics[width=0.32\textwidth]{figs/additional/brain_8x_file_brain_AXFLAIR_201_6002990_gt.png}
\end{subfigure}
\caption[*]{Brain images showing ground truth (left), reconstruction (middle) and reconstruction with added noise (right) at $8\times$ acceleration}
\label{fig:brain8x}
\end{figure}

\begin{figure}[ht]
\begin{subfigure}{\textwidth}
    \centering
    \includegraphics[width=0.32\textwidth]{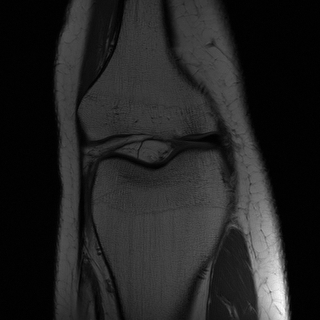}
    \includegraphics[width=0.32\textwidth]{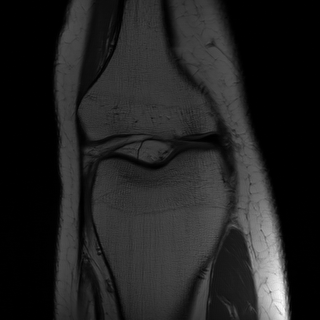}
    \includegraphics[width=0.32\textwidth]{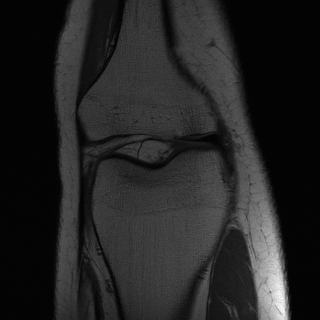}
\end{subfigure}

\vspace{5 mm}

\begin{subfigure}{\textwidth}
    \centering
    \includegraphics[width=0.32\textwidth]{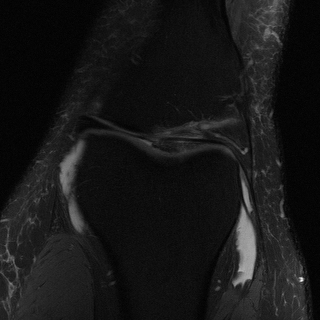}
    \includegraphics[width=0.32\textwidth]{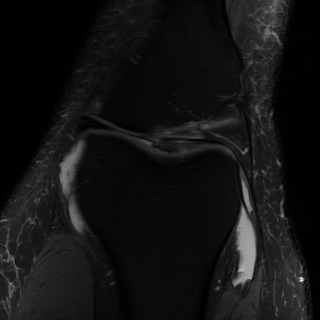}
    \includegraphics[width=0.32\textwidth]{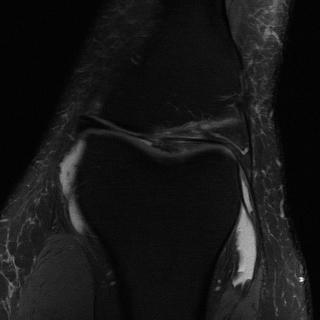}
\end{subfigure}
\caption[*]{Knee images showing ground truth (left), reconstruction (middle) and reconstruction with added noise (right) at $4\times$ acceleration}
\label{fig:knee4x}
\end{figure}

\begin{figure}[ht]
\begin{subfigure}{\textwidth}
    \centering
    \includegraphics[width=0.32\textwidth]{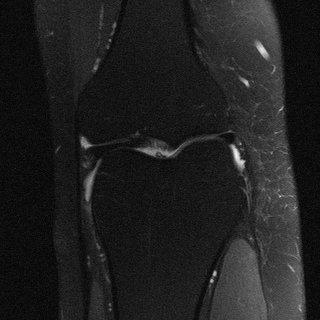}
    \includegraphics[width=0.32\textwidth]{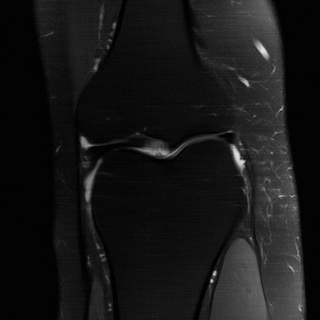}
    \includegraphics[width=0.32\textwidth]{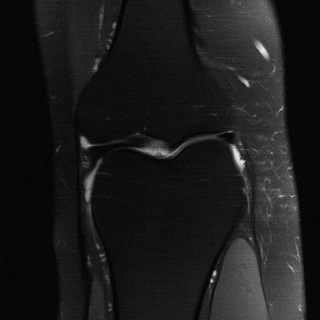}
\end{subfigure}

\vspace{5 mm}

\begin{subfigure}{\textwidth}
    \centering
    \includegraphics[width=0.32\textwidth]{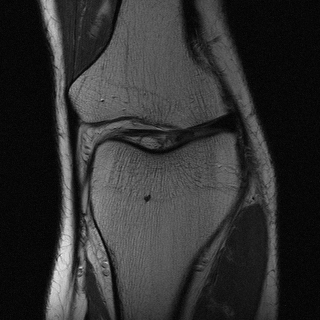}        \includegraphics[width=0.32\textwidth]{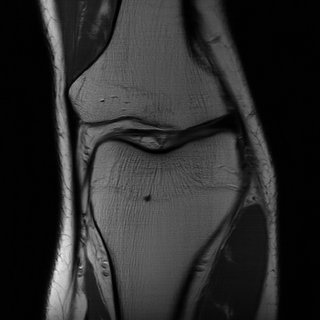}        \includegraphics[width=0.32\textwidth]{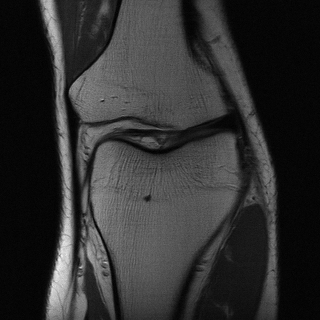}
\end{subfigure}
\caption[*]{Knee images showing ground truth (left), reconstruction (middle) and reconstruction with added noise (right) at $8\times$ acceleration}
\label{fig:knee8x}
\end{figure}





\end{document}